\documentclass[
reprint,
superscriptaddress,
amsmath,
amssymb,
pra,
floatfix,
]{revtex4-1}
\usepackage{amsfonts,amsmath,amssymb,amsthm}
 \usepackage{array,bm,color}
\usepackage{epsfig,graphicx,nomencl,revsymb4-1,upgreek,url}
\usepackage{hyperref}
\hypersetup{
    colorlinks=true,
    linkcolor=magenta,
    filecolor=black,      
    urlcolor=magenta,
    citecolor=magenta
}
\usepackage{algorithm}
\usepackage{algpseudocode}
\usepackage{graphicx}
\usepackage{calc}
\usepackage{siunitx}
\usepackage{tabularx}
\newcolumntype{Y}{>{\centering\arraybackslash}X}
\graphicspath{{./figures/}}

\newcommand{\bramatket}[3]{\left\langle #1 | #2 | #3 \right\rangle}

\newcommand{\note}[1]{}

\usepackage[utf8]{inputenc}

\begin{document}
\title{The QAOA with Few Measurements}

\author{Anthony M. Polloreno}
\email[Email: ]{ampolloreno@gmail.com}
\affiliation{JILA and University of Colorado, Boulder, Colorado 80309, USA}
\author{Graeme Smith}
\affiliation{JILA and University of Colorado, Boulder, Colorado 80309, USA}

\date{\today}

\begin{abstract}
\noindent The Quantum Approximate Optimization Algorithm (QAOA) was originally developed to solve combinatorial optimization problems, but has become a standard for assessing the performance of quantum computers. Fully descriptive benchmarking techniques are often prohibitively expensive for large numbers of qubits ($n \gtrsim 10$), so the QAOA often serves in practice as a computational benchmark. The QAOA involves a classical optimization subroutine that attempts to find optimal parameters for a quantum subroutine. Unfortunately, many optimizers used for the QAOA require many shots ($N \gtrsim 1000$) per point in parameter space to get a reliable estimate of the energy being minimized. However, some experimental quantum computing platforms such as neutral atom quantum computers have slow repetition rates, placing unique requirements on the classical optimization subroutine used in the QAOA in these systems. In this paper we investigate the performance of two choices of gradient-free classical optimizer for the QAOA - dual annealing and natural evolution strategies - and demonstrate that optimization is possible even with $N=1$ and $n=16$. 
\end{abstract}


\maketitle

\section{Introduction}
The Quantum Approximate Optimization Algorithm (QAOA) is a hybrid quantum-classical algorithm that uses a subroutine run on a quantum computer together with a classical optimizer to find approximate solutions to combinatorial optimization problems \cite{farhi2014quantum, hadfield2019quantum}. While many results have demonstrated that the QAOA faces substantial challenges if it is to outperform classical algorithms \cite{wurtz2020maxcut,gonzalez2022error,bravyi2020obstacles,farhi2020quantum,bittel2021training,hastings2019classical,hastings2021classical,akshay2020reachability, guerreschi2019qaoa}, it has nevertheless become a standard technique for benchmarking quantum computers \cite{harrigan2021quantum, otterbach2017unsupervised, graham2021demonstration}, serving as a holistic test. 

The classical optimization routine in the QAOA is used to select a collection of angles that parameterize a quantum circuit. While \cite{farhi2014quantum} discusses the feasibility of an open-loop grid search for the QAOA, in practice classical optimizers are closed-loop and require accurate estimates of the energy expectation value for any choice of angles. Previous work \cite{harrigan2021quantum, otterbach2017unsupervised, pagano2020quantum, hodson2019portfolio} has often chosen classical optimizers that require a large number of measurement repetition rates \cite{harrigan2021quantum, farhi2014quantum, zhou2020quantum, shaydulin2019evaluating, alexeev2020reinforcement, shaydulin2019multistart}. Furthermore, theoretical works also often assume that the optimizer has access to the expectation value at each point in parameter space \cite{streif2021beating, tate2020bridging}.

However, this assumption is not always justified - some quantum computing platforms, such as neutral atoms, have substantially slower circuit repetition rates. In the case of neutral atoms, this is due in part to atom reloading \cite{baker2021exploiting, saffman2016quantum}. Modern neutral atom quantum computers use lossy measurements which remove atoms from the trap, requiring atoms to be reloaded between measurements. Moreover modern systems such as \cite{graham2021demonstration}, can have measurement durations as long as $30$ms. These facts together can give total shot-to-shot measurement rates of around $5$Hz \cite{saffman2022measurementrate}.

This places substantial difficulties on the classical optimizer used in the QAOA. While noise is commonly modeled as coming from decoherence and miscalibration, it can also be a result of shot noise in sampling from the quantum computer a limited number of times. This noise can cause unreliable estimates of quantities such as the gradient and the energy. 

In this paper we explore a classical optimization routine for circumventing these difficulties -- dual annealing. Dual annealing is a form of simulated annealing, which explores the search space while annealing a ``temperature'' that parameterizes how likely the algorithm is to jump to higher energy configurations. This can be useful for avoiding local minima. We first discuss modifications that can be made to the algorithm to accommodate the error from taking a limited number of shots. Next, we find empirically that this optimizer performs well even without modification. In particular, we find that dual annealing can find global maxima with a single shot at each point in parameter space that it evaluates.

In addition to dual annealing, we consider natural evolution strategies (NES). NES refers to a family of recent optimization routines that sample from a parameterized distribution on the search space. By evolving this distribution based on values measured, NES can search for global maxima in a way that is robust against local optima. As we will see, it also has a built-in robustness against noisy evaluations of the cost function that arise from small sample sizes. Previous work \cite{anand2021natural} has studied the use of NES in the QAOA. In this work, we focus specifically on the ability of NES to perform despite small sample sizes, for up to $n=16$ qubits.

To assess the efficacy of this optimization technique, we propose a family of graphs and use the QAOA to find the maximum cut of a member of this family. We provide complexity theoretic arguments that suggest these graphs have maximum cuts that may be hard for classical computers to approximate. While several families of graphs are known to be poor choices for showing quantum advantage \cite{wurtz2020maxcut, farhi2020quantum, farhi2014quantum, 2005.08747}, it is an open problem to find families of graphs on which the QAOA can be shown to outperform classical algorithms.  

\section{Theory}\label{sec:theory}
\begin{figure*}
    \centering
    \includegraphics[scale=1]{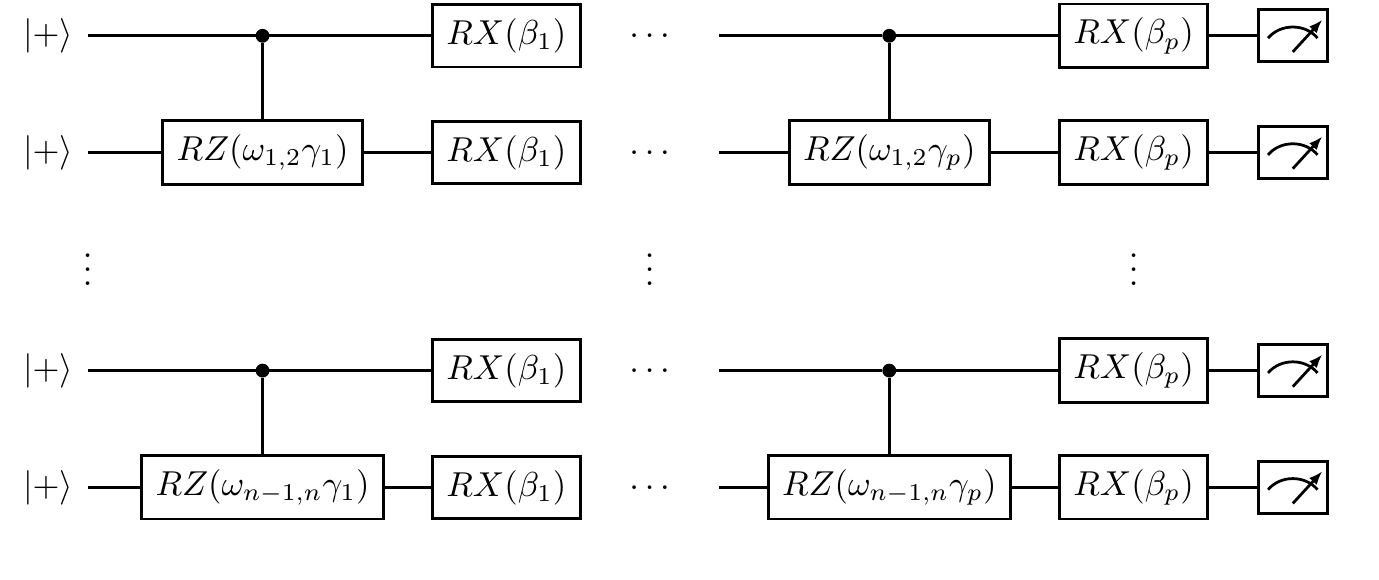}
    \caption{An example of a circuit implementing the QAOA from Eq.~\ref{eq:QAOA}, using the MAX-CUT Hamiltonian in Eq.~\ref{eq:hamiltonian}. In this example, we've assumed that there are only edges between nearest neighbors.}
    \label{fig:qaoa_circuit}
\end{figure*}

The quantum circuit used by the QAOA is described by $2p$ parameters, in a layered ansatz. Specifically, the unitary evolution of the quantum computer has input angles $\vec{\gamma}, \vec{\beta}$ and is given as
\begin{eqnarray}\label{eq:QAOA}
    U(\vec{\gamma}, \vec{\beta}) = \prod_{j=1}^{p}e^{-i\beta_jB}e^{-i\gamma_j H},
\end{eqnarray}
where $H$ is called the driver and  $B = \sum_{k=1}^n\sigma^x_k$ is called the mixer on $n$ qubits. An example of this circuit is shown in Fig.~\ref{fig:qaoa_circuit} and the details of $H$ for the problem considered in this paper are given in Eq.~\ref{eq:hamiltonian}. $H$ is a Hamiltonian whose ground state encodes the solution an optimization problem. In this paper we consider the MAX-CUT problem, with associated Hamiltonian
\begin{equation}\label{eq:hamiltonian}
    H  = \sum_{(i, j)\in E} \frac{1}{2}\omega_{ij}(1 + \hat{Z}_i\hat{Z}_j).
\end{equation}
Here $E$ is the set of edges in a graph, and $\omega_{ij}$ are the weights for the associated edges.

The goal of the QAOA is to find optimal values of $\vec{\gamma}$ and $\vec{\beta}$ that minimize
\begin{equation}\label{eq:cost}
C(\vec{\gamma}, \vec{\beta}) = \bramatket{+}{U(\vec{\gamma}, \vec{\beta})^{\dagger}HU(\vec{\gamma}, \vec{\beta})}{+}.
\end{equation}
In this paper, instead of minimizing $C(\vec{\gamma}, \vec{\beta})$ we equivalently maximize $-C(\vec{\gamma}, \vec{\beta})$. We will only consider the $p=1$ QAOA, and therefore only have $\gamma_1$ and $\beta_1$. Because there is only one of each angle in the examples we consider, we will refer to them as $\gamma$ and $\beta$ for the rest of the paper. Numerous descriptions of the QAOA and MAX-CUT exist in the literature \cite{farhi2014quantum, otterbach2017unsupervised}, and further discussion is omitted here. MAX-CUT is a common problem to consider for demonstrations of QAOA due to the locality of the Hamiltonian, requiring only two-body terms. This makes it attractive for early demonstrations of hardware such as that in \cite{graham2021demonstration}.

To find the optimal angles, $\gamma$ and $\beta$, we consider simulated annealing algorithms. Simulated annealing is traditionally introduced in analogy to the metallurgical process, wherein the optimizer has a temperature which controls random fluctuations and an energy function that is used for preferentially exploring certain configurations. The system has ``thermal kinetic energy'', with a visiting temperature which allows it to probabilistically explore the search space while avoiding local minima. The transition between any two points has an associated hopping probability that characterizes how likely a transition between the two states is. As the temperature is lowered, exploration becomes more difficult in the presence of energy barriers.

Simulated annealing over a discrete set (such as that achieved by discretizing $\gamma$ and $\beta$) can be shown to converge to the set of global minima \cite{bertsimas1993simulated}. For each temperature, the hopping probabilities form a Markov chain, whose stationary distribution is given by 
\begin{equation}
    \pi_T(i) = \frac{1}{Z_T}\exp{(-\frac{J(i)}{T})}
\end{equation}
\cite{bertsimas1993simulated}, with $Z_T$ a normalizing constant. In the limit that $T\rightarrow 0$, we see that this distribution concentrates around the collection of minima of $J$.

We further note that if the energy estimate is noisy, this convergence is not guaranteed. This can easily be seen by considering the space with two elements, $i_1$ and $i_2$, and function $K$ such that $K(i_1) = 0$ and
\begin{equation}
    K(i_2)= 
\begin{cases}
    2,& \text{with probability } 1/2\\
    -1,              & \text{otherwise}
\end{cases}.
\end{equation}
Taking $J = \langle K \rangle$ to be the energy function, we see that minimum is $i_1$, however the stationary point of the Markov chain will be the uniform distribution over $i_1$, $i_2$.

The algorithm can, however, be modified to maintain its convergence properties. We can model finite shot noise as a random variable for the $k^{th}$ step in the annealing chain that depends on the state being measured,
\begin{equation}
X_k = (\frac{1}{N}\sum^N_{\ell=1} f_\ell(x_j) - J(x_j)) - (\frac{1}{N}\sum^N_{\ell=1} f_\ell(x_i) - J(x_i)),
\end{equation}
 where $J(m) = \bramatket{m}{C}{m}$ and $f_\ell$ is the $\ell^{th}$ evaluation on the quantum computer - i.e. the $\ell^{th}$ shot. One of the conditions for convergence given in \cite{gelfand1989simulated} is that $\sigma_k \in o(T_k)$, where $\sigma_k$ is the standard deviation of $X_k$, $T_k$ is the annealing temperature and $o(\cdot)$ is the standard asymptotic notation for a strict upper bound. Likewise in what follows, $O(\cdot)$ denotes an asymptotic upper bound with possible equality, $\omega(\cdot)$ denotes a strict asymptotic lower bound, $\Omega(\cdot)$ denotes an asymptotic lower bound with possible equality and $\Theta(\cdot)$ means both $O(\cdot)$ and $\Omega(\cdot)$.

From \cite{farhi2014quantum} we know that the values of the sampled cost function are concentrated about their mean $C(\gamma, \beta)$, with variance, $\sigma^2$ upper bounded by
\begin{equation}
    \sigma^2 \leq 2e\frac{(v-1)^{(2p+2)} - 1}{(v-1) - 1},
 \end{equation}
 for a graph with $e$ edges and $v$ vertices. Thus
 \begin{equation}\label{eq:variance_bound}
     \sigma^2_{\hat{\mu}} = \sigma^2/N \in O(e/N),
 \end{equation}
where $\sigma^2_{\hat{\mu}}$ is the variance of sample mean. This shows that we should expect the number of shots required to be at worst proportional to the number of edges in the graph.

What we will see next is that optimization is possible even with $N=1$. The particular graph considered in this section is shown in Fig.~\ref{fig:graph}, and is sampled from the distribution described in Sec.~\ref{sec:graph}. The graph is chosen from a distribution with $e = \frac{3v}{5}$ edges, with $v=20$, and therefore only contains $16$ vertices with edges. Such a graph would be useful for demonstrating the QAOA on an architecture of a $4\times 4$ grid of atoms. Using Eq.~\ref{eq:variance_bound}, with N=1, we find an extremely loose bound on the sample variance of $86,786$. Despite this, we will see that both optimizers, dual annealing and NES, perform well.

From \cite{gelfand1989simulated}, we expect the modified annealing algorithm to work if $\sqrt{N/e} \in \omega(1/T_k)$. We will show, however, in Sec.~\ref{sec:numerics} that even without modifying dual annealing, it can find the local maxima with a single shot per point in parameter space. 

The implementation details of the dual annealing algorithm can be found in \cite{2020SciPy-NMeth} and theoretical details can be found in \cite{xiang1997generalized}. We reproduce some relevant details here. In particular, the algorithm uses a distorted Cauchy-Lorentz distribution for the ``visiting temperature''

\begin{equation}
    g_{q_v}(\Delta x(t)) \propto \frac{[T_{q_v}(t)]^{-\frac{D}{3-q_v}}}{[1 + (q_v - 1)\frac{(\Delta x(t))^2}{[T_{q_v(t)}]^{\frac{2}{3-q_v}}}]^{\frac{1}{q_v-1} + \frac{D-1}{2}}},
\end{equation}
where $x(t)$ is a jumping distance, and $D$ is the dimension of the optimization problem \cite{xiang1997generalized}. For the $p=1$ QAOA $D=2$, since there are only two values to optimize, $\gamma$ and $\beta$. This temperature is annealed according to 
\begin{equation}
    T_{q_v}(t) = T_{q_v}(1)\frac{2^{q_v - 1} - 1}{(1 + t)^{q_v - 1} - 1},
\end{equation}
and new points are accepted with probability
\begin{equation}
    p_{q_a} = {\rm min}\{1, [(1 - q_a)\Delta E/T_{q_v}]^{\frac{1}{1-q_a}}\}. 
\end{equation}

$q_a$ and $q_v$ are called the acceptance and visiting parameters, and are hyperparameters used to control how exploratory the annealing algorithm is. This annealing routine is accompanied by local search at the end of each annealing attempt and a number of restart attempts. Simulated annealing, and hence dual annealing, benefits from a wealth of analysis and convergence guarantees \cite{hajek1988cooling}. 

The other family of optimizers we consider are natural evolution strategies (NES). NES are a family of optimization algorithms that take inspiration from biology. By evaluating a family of points $\mathcal{P}$, NES can estimate a gradient, as was shown in \cite{wierstra2008natural}. Namely, \cite{wierstra2008natural} shows that, for a Gaussian distributed family,
\begin{eqnarray}
    \nabla_{\mu} \mathbb{E}(f(x)) \approx \frac{1}{M} \sum_{k=1}^{M} f(z_k) \nabla_{\mu}\log{\pi(z_k|\theta)},\\
    \nabla_{\mu} \mathbb{E}(f(x)) \approx \frac{1}{M} \sum_{k=1}^{M} \frac{1}{\sigma}f(z_k)(z-\mu)\\
    \nabla_{\mu} \mathbb{E}(f(x)) \approx \frac{1}{M} \sum_{k=1}^{M} \frac{1}{\sigma}f(z_k)Z
\end{eqnarray}

$Z = z-\mu$ is a unit variance, origin centered random variable, and $z_k$ are sampled from the Gaussian family. Furthermore, the function $f$ itself arises as an expectation value, from sampling the state $\rho$ on the quantum computer, giving 
\begin{equation}
    \nabla_{\mu} \mathbb{E}(f(x)) \approx \frac{1}{M} \sum_{k=1}^{M} \frac{1}{\sigma}\mathbb{E}_{\rho}(f(\mu + \sigma Z)Z.
\end{equation}

For a population with $M$ members, and using $N$ shots per member, we therefore find the following estimator:
\begin{align}\label{eq:good_estimator}
        \nabla_\mu \mathbb{E}(f(x)) = \frac{1}{\sigma NM}\sum^M_i\sum^N_j f_j(x_i)Z_i,
\end{align}
where $f_j$ gives the $j^{\rm th}$ circuit execution on the quantum computer, and $x_i$ is the $i^{\rm th}$ member of the population.

Typically if the number of shots $N$ is small, the conventional estimator $\hat{f}(x) = \frac{1}{N}\sum_jf_j(x)$ is poor, however by exchanging the sums in Eq.~\ref{eq:good_estimator} we see that as long as the function does not vary drastically over the population, we obtain a good estimate of the gradient despite taking a relatively small number of shots. We also explore more of the search space than if we had just evaluated many shots at a single point. Ref. \cite{anand2021natural} contains a more detailed exploration of the efficacy of NES in general variational quantum circuits - this paper will focus solely on their utility in the low shot count regime. 

\section{Choice of Graph}\label{sec:graph}
While the primary purpose of this paper is to discuss the choice of classical optimizer in the QAOA, we need to choose a problem instance. So as to avoid solving a problem that is known to be trivial, we give a heuristic argument for our choice of graph. The particular problem we choose due to its simple-to-implement Hamiltonian and abundance in the quantum computing community is MAX-CUT. The difficulty of MAX-CUT depends on the family of graphs being considered, and the general problem has been well-studied classically - the Goemans-Williamson (GW) algorithm gives an optimal approximation ratio assuming the unique games conjecture \cite{khot2007optimal, mossel2005noise, trevisan2012khot}. Therefore one would like to identify a collection of graphs that classical computers perform worst on.

In this work we choose random graphs $G(v,e)$ on $v$ vertices with $e$ edges chosen at random from all possible edges. MAX-CUT on these graphs has a phase transition when $e/v=1/2$, in the following sense: below $1/2$ there are efficient classical algorithms for solving the problem, however above $1/2$ there seem to be hard instances \cite{coppersmith2004random, gamarnik2018max}. Furthermore, if we consider signed graphs, then if the cover number of the positively signed edges is near $\sqrt{v}$, the problem becomes strongly NP hard \cite{garey1978strong}. A problem is strongly NP hard if the problem remains NP hard when its numerical parameters are bounded by a polynomial in the input size. This means that hard instances are, roughly speaking, easy to generate, and so in particular randomly generating instances should result in graphs that may be more challenging for classical algorithms.

\begin{figure}
  \centering
  \includegraphics[width=\columnwidth]{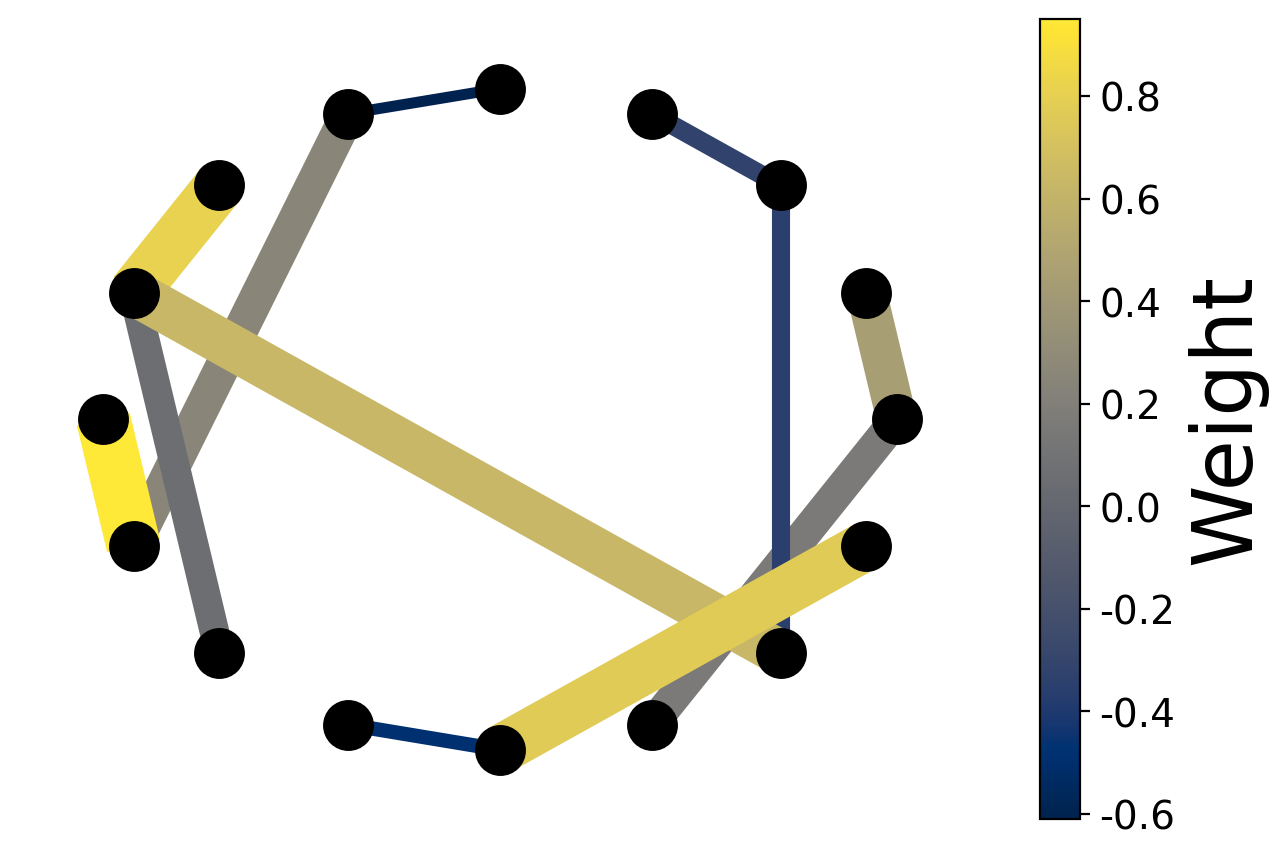}
  \caption{\textbf{Example Graph from Section~\ref{sec:graph}.} Above we see a graph on $v=20$ vertices, generated according to the distribution described in Sec.~\ref{sec:graph}.  Edgeless vertices are omitted from the figure. We sample $e = 3v/5$ edges at random, and edge weights are chosen uniformly at random from $[-1, 1]$.}
  \label{fig:graph}
\end{figure}

The cover number of the positively signed edges is the minimum size of a subset $V$ of the vertices, such that every edge with positive weight has at least one endpoint in $V$. In particular, \cite{mccormick2003easy} shows this family is strongly NP hard for graphs with cover number $\Omega(v^{-k})$, with $k$ a positive integer. We have already specified $e\in \Theta(v)$ so that $k=2$ is the smallest $k$ that is guaranteed to be compatible. This suggests that algorithms like the GW algorithm may perform less favorably on these graphs. An example of this graph is shown in Fig.~\ref{fig:graph}. Similar graphs have been considered in \cite{tate2020bridging}.

\section{Numerics}\label{sec:numerics} 
We simulate the $p=1$ QAOA on a graph sampled from the family of graphs discussed in Sec.~\ref{sec:graph} using quimb \cite{gray2018quimb}. We emphasize that the purpose of this paper is to demonstrate the performance of the optimizers with low shot rates. The $p=1$ QAOA has been well-studied, and its performance is well-understood analytically \cite{wang2018quantum,wurtz2021maxcut, farhi2014quantum}. Thus, our emphasis is that these techniques allow for the demonstration and benchmarking of a quantum computer - for the case of $p=1$ optimization, one should use the known-optimal choice of angles if the goal is produce a best approximate solution to the problem being studied. In certain problem instances of MAX-CUT, we can reduce the search space for both $\gamma$ and $\beta$ to the range $[0, \pi]$ \cite{zhou2020quantum}. In general we see this is not possible, since scaling the edge weight by a factor of $1/c$ moves the maximum in $\gamma$ by a factor of $c$. Nevertheless, we restrict our search space, since we are interested primarily showing that optimization is possible. For our example, we choose a graph on $v=20$ vertices, which has $n=16$ qubits that interact with at least one other qubit. We note that $n=16$ is a non-trivial example for initial demonstrations of new quantum computing architectures \cite{graham2021demonstration}.

The particular graph considered in this section is shown in Fig.~\ref{fig:graph}, and is sampled from the distribution described in Sec.~\ref{sec:graph}. The graph is chosen from a distribution with $e = \frac{3v}{5}$ edges, with $v=20$, and therefore only contains $16$ vertices with edges. Such a graph would be useful for demonstrating the QAOA on an architecture of a $4\times 4$ grid of atoms. As mentioned in Sec.~\ref{sec:theory}, rather than minimizing Eq.~\ref{eq:cost}, we maximize $-C(\gamma, \beta)$.

\begin{figure}
  \centering
  \includegraphics[width=\columnwidth]{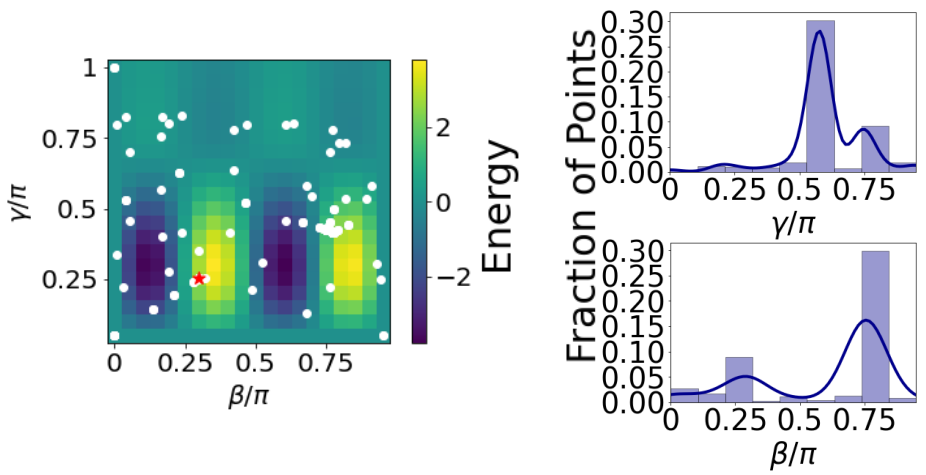}
  \caption{\textbf{Optimization using dual annealing.} Each white point corresponds to a single shot measured by dual annealing. The kernel density estimates show peaks corresponding to the local maxima, demonstrating that optimizer preferentially samples from higher energies. Moreover, the red star shows the final sampled point which lies near a global maximum.}
  \label{fig:annealing}
\end{figure}

In Fig.~\ref{fig:annealing} we see the performance of dual annealing on the graph in Fig.~\ref{fig:graph}. We use one shot per point in parameter space ($N=1$) and allow the dual annealing algorithm ten opportunities to restart and re-anneal. In total the optimizer sampled $326$ times. For a clock rate of $5$Hz this would take just over a minute to run. Fig.~\ref{fig:annealing} also provides two kernel density estimates of the distribution of sampled points which highlights that there are two distinct attractors over $\beta$, and one over $\gamma$, corresponding to the optima of the landscape. Kernel density estimates use Gaussian kernels to produce a non-parametric estimate of the sampling distribution that is given by
\begin{equation}
    \hat{f}_h(x) = \frac{1}{nh}\sum \exp{(-(x-x_i)^2/h^2)},
\end{equation}
where the bandwidth $h$ given by 
\begin{equation}
    h = n^{-1/(d+4)},    
\end{equation}
and $n$ is the number of samples ($326$) and $d$ is the dimension ($1$) as suggested in \cite{scott2015multivariate}.
This demonstrates that the dual annealing optimizer not only can find a good cut, but also is taking advantage of the structure of the optimization landscape.

\begin{figure}
  \centering
  \includegraphics[width=\columnwidth]{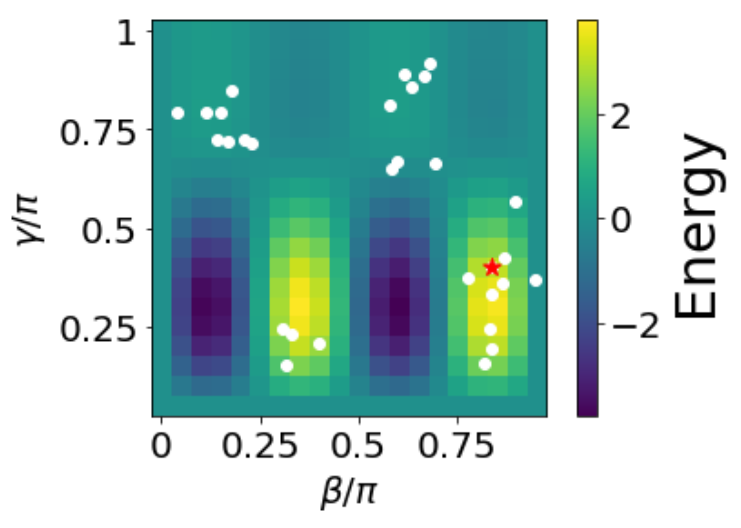}
  \caption{\textbf{Optimization using evolution strategies.} Each white point corresponds to the mean of a Gaussian distributed sample of ten points. The red star corresponds to the last sampled parameters after $30$ generations of evolution. The NES algorithm in general has explored all local maxima, and ended in one of the global maxima, with the final point labeled by a red star.}
  \label{fig:evolution}
\end{figure}

In Fig.~\ref{fig:evolution} we see that the evolution strategies optimizer is largely constrained to the four local maxima, one of which is the global maximum. In this example each call to the quantum computer evaluates only a single shot, $N=1$, and each population has $M=10$ members. This is evolved over $|\mathcal{P}| = 30$ generations, for a total of $300$ evaluations on the quantum computer. For a measurement repetition rate of $5$Hz this would take a minute to run.

We see that in both cases, dual annealing and natural evolution strategies, the optimizer is able to perform with just a single shot per choice of angle pair $\gamma$ and $\beta$. For quantum computers with low shot rates this provides a tractable method for performing closed-loop quantum-classical optimization, with noisy estimates of the energy at each point. 

\section{Conclusion}
Due to its easy implementation and interpretation, the QAOA is an appealing holistic benchmark for quantum computers. However, different quantum computing architectures place different requirements on the classical optimization subroutine of the QAOA. In particular, neutral atom quantum computers often have substantially slower measurement rates than conventional superconducting quantum computers. To use the QAOA as a benchmark for these platforms thus requires choosing a classical optimizer that can use noisy estimates of the QAOA energy.

In this paper, we have demonstrated two optimization techniques that are well suited for this problem, dual annealing and natural evolution strategies. Dual annealing searches the space randomly, accepting transitions if the cost is lower, or if the annealing temperature is sufficiently high to allow jumps to worse points. Natural evolution strategies is a technique from machine learning that evolves a Gaussian distributed population of points in parameter space which can avoid local minima. In both cases, the algorithms can be tuned to use a variable number of shots, down to a single shot per point, while still finding good cut values for the MAX-CUT problem on as many as $n=16$ qubits.

Additionally, we have suggested a new family of graphs which have a phase transition in their known classical hardness. There are several straightforward analyses to extend this work - including realistic noise models, scaling to larger numbers of qubits and extending the algorithm considered here to model the shot uncertainty explicitly \cite{gelfand1989simulated}, as described in Sec.~\ref{sec:theory}. Finally, while the intention of this work was to discuss the optimizer choice for demonstrating the performance of NISQ hardware, and thus focused on $p=1$, exploring the increased dimensionality of the search space for higher $p$ is an interesting direction for future research.

\section{Acknowledgements}
AMP thanks Danylo Lykov for suggesting quimb and Hari Krovi for suggesting the particular choice of graph. AMP also thanks Mark Saffman for providing details about the experiment in \cite{graham2021demonstration}. This material is based upon work supported by the Defense Advanced Research Projects Agency (DARPA) under Contract No. HR001120C0068. AMP acknowledges funding from a NASA Space Technology Graduate Research Opportunity award.

\bibliography{refs}
\end{document}